\documentclass[a4paper,11pt]{article}
\usepackage{pos}

\usepackage{braket}
\usepackage[utf8]{inputenc}
\usepackage{graphicx}    
\usepackage{subcaption}  

\title{Entanglement entropy by tensor renormalization group approach}
\ShortTitle{Entanglement entropy by tensor renormalization group approach}

\author[a]{Takahiro H{ayazaki}}
\author[b]{Daisuke K{adoh}}
\author[a]{Shinji T{akeda}}
\author*[b]{Gota T{anaka}}

\affiliation[a]{Institute for Theoretical Physics, Kanazawa University,\\
Ishikawa 920-1192, Japan}

\affiliation[b]{Institute for Mathematical Informatics, Meiji Gakuin University,\\
Kanagawa 244-8539, Japan}

\emailAdd{t\_hayazaki@hep.s.kanazawa-u.ac.jp}
\emailAdd{kadoh@mi.meijigakuin.ac.jp}
\emailAdd{takeda@hep.s.kanazawa-u.ac.jp}
\emailAdd{gotanak@mi.meijigakuin.ac.jp}

\abstract{
We report on tensor renormalization group calculations of entanglement entropy
in one-dimensional quantum systems.
The reduced density matrix of a Gibbs state can be represented as a $1 + 1$-dimensional 
tensor network, which is analogous to the tensor network representation of the partition function.
The HOTRG method is used to approximate the reduced density matrix for arbitrary subsystem sizes,
from which we obtain the entanglement entropy.
We test our method in the quantum Ising model and obtain the entanglement entropy of the ground state
by taking the size of time direction to infinity.
The central charge $c$ is obtained as $c = 0.49997(8)$ for a bond dimension $D=96$,
which agrees with the theoretical value $c=1/2$ within the error.  
}

\FullConference{%
  The 41st International Symposium on Lattice Field Theory (LATTICE2024)\\
  28 July - 3 August 2024\\
  Liverpool, United Kingdom
}

\begin{document}
\maketitle

\section{Introduction}
Quantum entanglement is a non-local correlation between two subsystems $A$ and $\bar A$ of a quantum many-body system,
which is quantified by entanglement entropy,
\begin{align}
    S_A = - \mathrm{Tr} ( \rho_A \log \rho_A),
        \label{eq.def_EE}
\end{align}
where $\rho_A=\text{Tr}_{\bar A} \rho$ is a reduced density matrix of subsystem $A$. 
Entanglement entropy plays a crucial role in various fields such as quantum field theory, quantum gravity, and information theory.

Although entanglement entropy can be obtained analytically in certain cases 
\cite{Calabrese:2004eu, Calabrese:2009qy, Ryu:2006ef},
establishing a general method to calculate this quantity
in interacting field theories remains a challenging task.
The Monte Carlo (MC) method is a typical way to approach the problem.
In fact, the entanglement entropy cannot be directly obtained because it is difficult to rewrite equation \eqref{eq.def_EE}
into a path integral form to which the MC method can be applied.
Instead, with so-called "replica trick", the MC method evaluates $s$-th Reny entropy 
from which the entanglement entropy is extracted using an additional extrapolation $s\rightarrow 1$ 
\cite{Velytsky:2008sv, Buividovich:2008kq, Nakagawa:2009jk,
Nakagawa:2010kjk,Itou:2015cyu,Rabenstein:2018bri,Bulgarelli:2023ofi,Jokela:2023rba}.
Since it is difficult to evaluate the systematic errors associated with this extrapolation and to approach systems
with sign problems for which MC methods are not well suited, it will be important to find another solution.

The tensor renormalization group (TRG) method is a promising numerical approach to quantum field theory.
The reduced density matrix is expressed as a tensor network, and the TRG can be used to evaluate entanglement entropy
\cite{ueda2014doubling, Yang:2015rra, Bazavov:2017hzi, Luo:2023ont}. Except for the case where $A$ is exactly half of the total space, no general TRG algorithm for subsystems of arbitrary size is yet known.

In this report, we propose a new algorithm based on the higher-order tensor renormalization groups algorithm \cite{Xie:2012mjn}
for computing the entanglement entropy of subsystems of arbitrary size.
We apply our algorithm to the one-dimensional quantum Ising model
at the critical point and evaluate the entanglement entropy of the ground state.
The obtained central charges agree very well with the theoretical value.

This report is organized as follows.
In Section 2, we explain how to compute the entanglement entropy
using the tensor renormalization group method along with the details of our method.
The numerical results are presented in Section 3, and
the last section is devoted to the summary and outlook.

\section{Theory}
\subsection{One-dimensional quantum Ising model}
We derive the tensor network representation of the reduced density matrix of a Gibbs state $\rho=e^{-\beta H}/\mathrm{Tr}(e^{-\beta H})$. 
First, the one-dimensional quantum system is represented as a $(1+1)$-dimensional classical system,
then the reduced density matrix is represented as a tensor network
in the same way that the partition function of the classical system is represented as such.

The Hamiltonian of the one-dimensional transverse field quantum Ising model is given by:
\begin{align}
    \hat{{H}} = - J \sum_{\braket{i,j}} \sigma^z_{i} \sigma^z_{j}
        - h_x \sum_{i} \sigma^x_i,  \label{eq.quantum_Hamiltonian}
\end{align}
where $i, j \in \{1,2, \dots, L \}$ label lattice sites, 
and $\braket{i,j}$ denotes all possible pairs of nearest neighbor lattice sites.
The periodic boundary condition is assumed.
Matrices $\sigma^x_i, \sigma^y_i$, and $\sigma_i^z$ are Pauli matrices defined at site $i$.
Pauli matrices at different sites commute with each other.

A state $\ket{ \{s\}}$ is defined by $\ket{ \{s\}} \equiv \bigotimes_i \ket{s_{i}}$,
where $\ket{s_{i}}$ is an eigenstate of $\sigma^z_i$
and satisfies $\sigma^z_i \ket{s_{i}} = s_{i} \ket{s_{i}}$, and $s_{i} = \pm 1$.
Using the Suzuki-Trotter decomposition
\begin{align}
    e^{-\beta \hat{H}} = \lim_{N\to\infty} \left( e^{\frac{\beta J}{N} \sum_{\braket{i,j} }\sigma^z_i \sigma^z_j} e^{\frac{\beta h_x}{N} \sum_i \sigma^x_i}\right)^N,
\end{align}
we calculate a matrix element $e^{-\beta \hat{H}}$.
Inserting $N-1$ complete systems, we obtain
\begin{align}
    \braket{\{s\}_N | e^{-\beta \hat{H}} | \{s\}_0}
    = \lim_{N\to\infty} \sum_{\{s\}_1, \{s\}_2, \dots, \{s\}_{N-1}}
    C \exp \left[ K \sum_{\braket{i,j}} \sum^{N-1}_{t=0} s_{i,t} s_{j,t}
        + K' \sum_i \sum_{t=0}^{N-1} s_{i,t} s_{i,t+1} \right],
        \label{eq.partition_function}
\end{align}
where $K=\frac{\beta J}{N}$, $K'=\frac{1}{2} \log \left( \coth (\frac{\beta h_x}{N}) \right)$,
and $C=\left( \frac{1}{2} \sinh \frac{2\beta h_x}{N} \right)^\frac{L}{2}$.
To keep $K$ and $K'$ finite, we need to take the limit $\beta\to\infty$.
Setting $\{ s\}_N = \{ s\}_0$ and taking the sum over $\{ s\}_0$,
$\mathrm{Tr}(e^{-\beta \hat H})$ is expressed as a partition function of (1+1)-dimensional classical Ising model with couplings $K,K'$.
For the isotropic case $K=K'$, the quantum critical point, $\beta=\infty$ and $J=h_x$, is equivalent to $K=1/2\log(1+\sqrt{2})$,
the critical point of the (1+1)-dimensional classical Ising model.

\begin{figure}[ht]
  \begin{minipage}[b]{0.48\linewidth}
    \centering
    \includegraphics[keepaspectratio, scale=0.55]{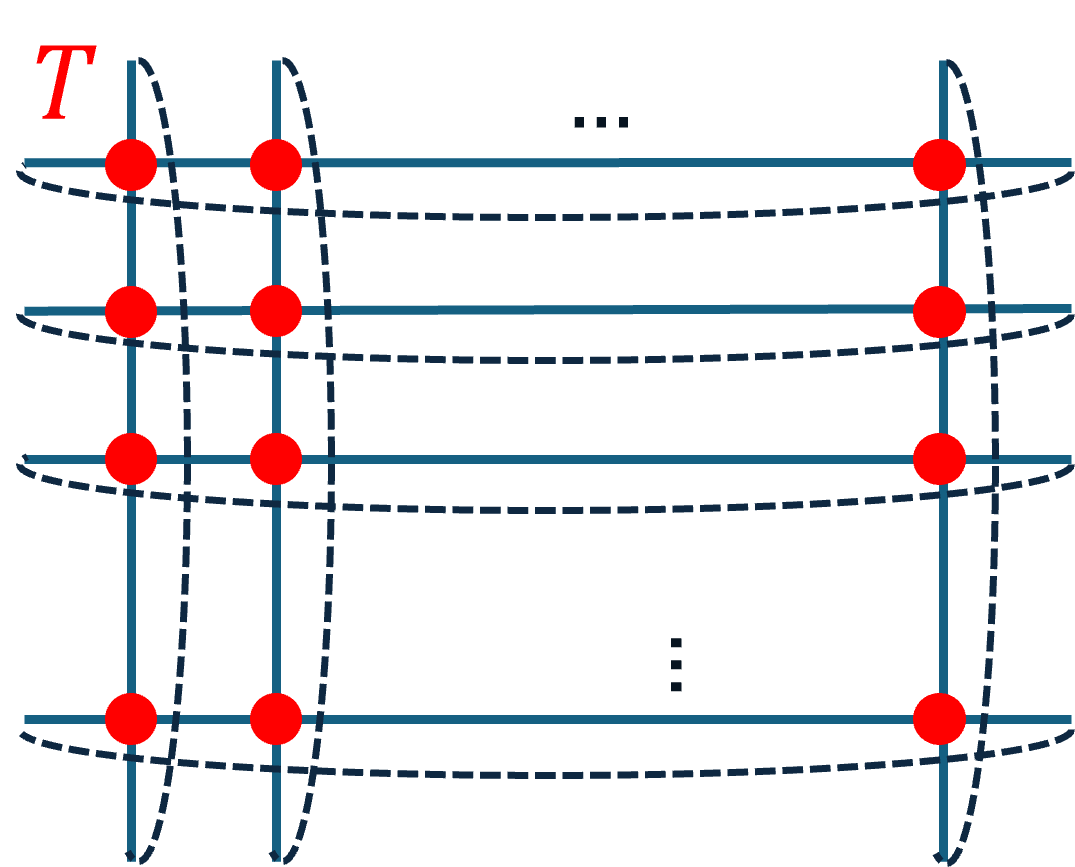}
    \subcaption{Partition function $Z$.}
    \label{fig.partition_function}
  \end{minipage}
    \begin{minipage}[b]{0.48\linewidth}
    \centering
    \includegraphics[keepaspectratio, scale=0.55]{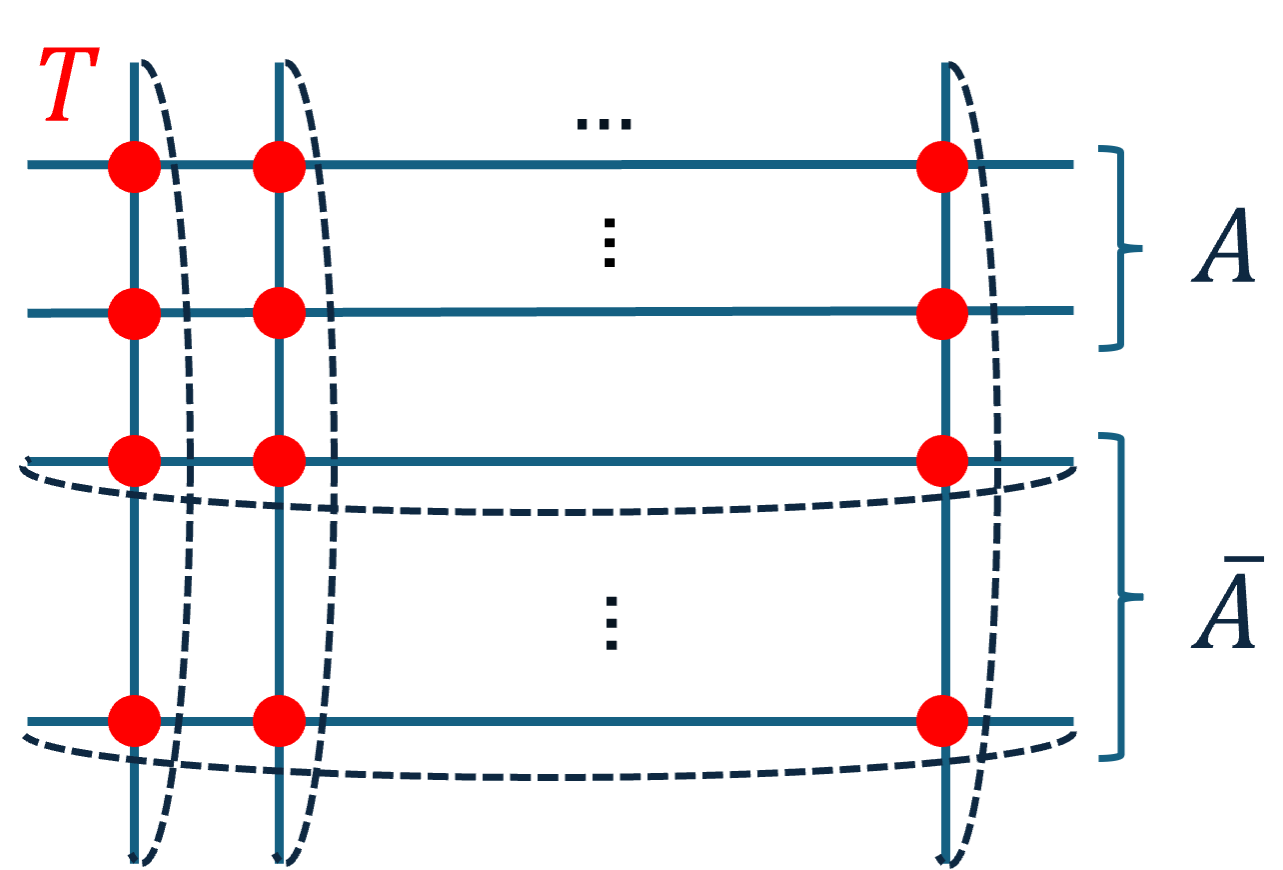}
    \subcaption{Reduced density matrix $\rho_A$.}
    \label{fig.reduced_density_matrix}
  \end{minipage}
  \caption{Tensor network representation of the partition function $Z$ and the reduced density matrix $\rho_A$.
    The horizontal axis represents the temporal direction, while the vertical axis represents the spatial direction.}
  \label{fig.tensor_networks}
\end{figure}

By choosing an appropriate basis $\ket{\{I\}}$, the density matrix can be expressed as a tensor network:
\begin{align}
    \braket{\{I\} | e^{-\beta \hat{H}} | \{I'\}} = \lim_{N\to \infty} (\mathcal{T}^N)_{\{I\}, \{I'\}},
    \label{eq.density_matrix_TN}
\end{align}
where $\mathcal{T}$ is a transfer matrix and is given by
\begin{align}
    (\mathcal{T})_{\{I\}, \{I'\}} = \sum_{a_i=1,2} T_{a_1 a_2 I_1 I'_1} T_{a_2 a_3 I_2 I'_2} \cdots T_{a_L a_{1} I_L I'_L},
    \label{eq.initial_tensor}
\end{align}
with a rank-4 tensor $T_{ijkl}$. See \cite{wareware_no_ronbun} for details on definitions.
Since $Z\equiv \mathrm{Tr}(e^{-\beta \hat H})=\mathrm{Tr}(\mathcal{T}^N)$, 
the partition function $Z$ can be expressed as a tensor network shown in Fig.~\ref{fig.partition_function}.
Since $\rho=\exp(-\beta \hat H)/Z$, the corresponding reduced density matrix is obtained by taking the partial trace of \eqref{eq.density_matrix_TN}. 
Thus, for $A=\{1,2,\cdots, l\}$, we find that the reduced density matrix of the Gibbs state is expressed as a tensor network
Fig.~\ref{fig.reduced_density_matrix}.
Note that the whole tensor indices are contracted for $Z$ while $l$ indices are open for $\rho_A$.

\subsection{Tensor renormalization group}
We show how to compute the entanglement entropy approximately through coarse-graining of the reduced density matrix
using the higher-order tensor renormalization group (HOTRG) algorithm\cite{Xie:2012mjn}.
The method considers the case where the subsystem $A$ is given by a single connected line segment,
but is applicable to any subsystem size.
In the following, let $L$ be the size of the spatial direction, 
$N$ the size of the time direction, and $l$ the size of the subsystem A.

\begin{figure}[ht]
    \centering
    \includegraphics[keepaspectratio, scale=0.6]{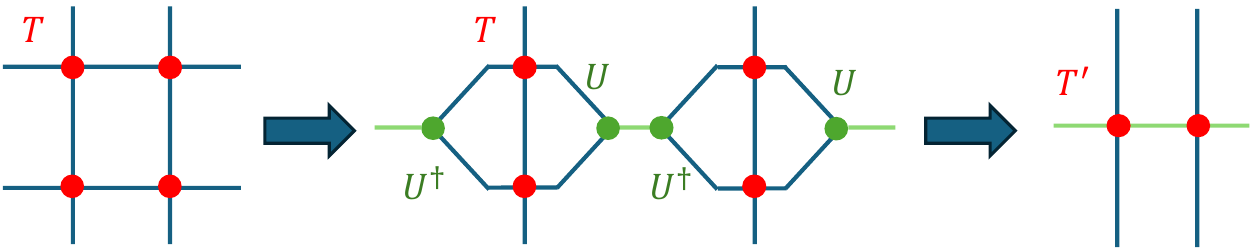}
    \caption{HOTRG procedure for the two-dimensional tensor network.}
    \label{fig.HOTRG}
\end{figure}

In the HOTRG algorithm, two neighboring tensors $T$ are coarse-grained into a single new tensor $T'$
as $T'=U^\dagger M U$, where $M=TT$.
The matrix $U$ is a $D^2 \times D_\text{cut}$ matrix that consists of eigenvectors of $M^\dagger M$ with $D_\text{cut}$ largest eigenvalues.
Since $U$ is a submatrix of a unitary matrix, it no longer satisfies $UU^\dagger = I$,
although it still satisfies $U^\dagger U = I$.
The HOTRG procedure is shown in Fig.~\ref{fig.HOTRG}.
The left figure in Fig. 2 shows a part of the network.
First, we insert $UU^\dagger$ into the network
as shown in the central figure in Fig.~\ref{fig.HOTRG}.
We then contract tensors and obtain $T'=U^\dagger M U$ as shown in the right figure in Fig.~\ref{fig.HOTRG}.
This procedure is applied alternately to each direction of the network,
and each application reduces the number of tensors by half.

We apply the HOTRG algorithm to $\rho_A$ to evaluate the entanglement entropy $S=-\mathrm{Tr}(\rho_A \log \rho_A)$.
Let $T^{(0)}$ be the initial tensor representing $\rho_A$. 
The spatial and temporal directions are renormalized alternately. These two renormalizations are counted as one set.
$T^{(k)}$ and $U^{(k-1)}$ are the renormalized tensor and the isometry that is required for the renormalization of spacial direction
at $k$ sets of renormalizations, respectively.
The isometries of the temporal direction are not needed for any purpose other than to create $T^{(k)}$.

\begin{figure}
    \centering
  \begin{minipage}[b]{0.48\linewidth}
    \centering
    \includegraphics[keepaspectratio, scale=0.5]{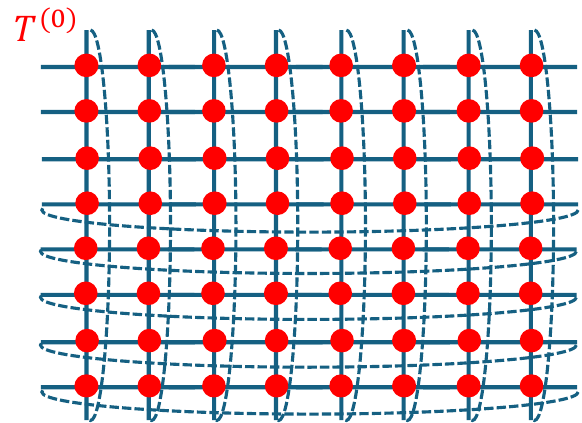}
    \subcaption{${\rho}_A$. }
    \label{fig.RDM_HOTRG_0}
  \end{minipage}
  \begin{minipage}[b]{0.48\linewidth}
    \centering
    \includegraphics[keepaspectratio, scale=0.5]{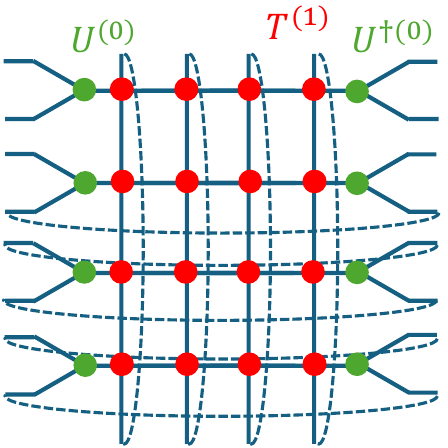}
    \subcaption{$\rho_A$ after a set of renormalization (n=1).}
    \label{fig.RDM_HOTRG_1}
  \end{minipage}
    \begin{minipage}[b]{0.48\linewidth}
    \centering
    \includegraphics[keepaspectratio, scale=0.5]{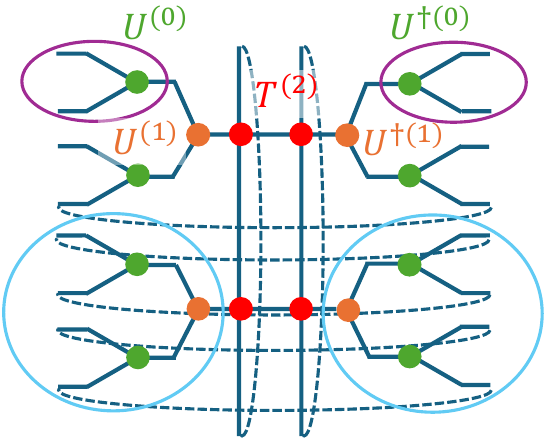}
    \subcaption{$\rho_A$ after two sets of renormalization (n=2).}
    \label{fig.RDM_HOTRG_2}
  \end{minipage}
  \begin{minipage}[b]{0.48\linewidth}
    \centering
    \includegraphics[keepaspectratio, scale=0.5]{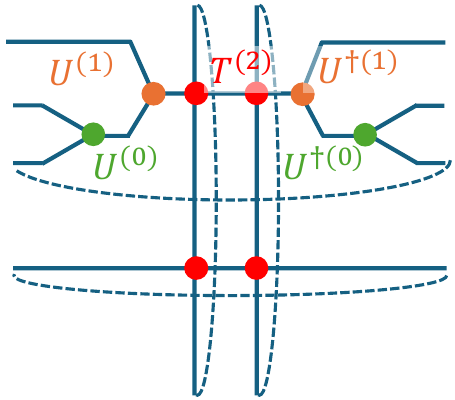}
    \subcaption{"Trimmed" $\rho_A$.}
    \label{fig.RDM_HOTRG_3}
  \end{minipage}
  \caption{Coarse-graining procedure for ${\rho}_A.$}
\end{figure}

To illustrate our method, we first consider the case of $L=N=8$ and $l=3$, as shown in Fig.~\ref{fig.RDM_HOTRG_0}.
In Fig \ref{fig.RDM_HOTRG_0}, note that six external lines are open, and $\rho_A$ is a $D^3 \times D^3$ matrix. 
Fig.~\ref{fig.RDM_HOTRG_1} and \ref{fig.RDM_HOTRG_2} denote $\rho_A$ after a set and two sets of renormalizations, respectively.
In the part excluding the boundary, the isometries $U^{(0)}, U^{(1)},\cdots U^{(k-1)}$s are used to create $T^{(k)}$, 
but on the boundary, they remain as tree graphs.

In addition, isometry tensors encircled by a purple line in Fig.~\ref{fig.RDM_HOTRG_2}  can be ignored
since they do not contribute to the entanglement entropy.
To see this, let $\rho'_A$ be the rest of the network.
The entanglement entropy $S_A$ satisfies
$S_A=-\mathrm{tr} {\rho}_A \log {\rho}_A=-\mathrm{tr} U {\rho}'_A U^\dagger \log (U {\rho}'_A U^\dagger)
=-\mathrm{tr} {\rho}'_A \log {\rho}'_A$, where we have used the property $U^\dagger U = I$.
The isometry tensors encircled by the blue line also do not contribute to $S_A$
because they coincide with an identity matrix in $\rho'_A$ using $U^\dagger U=I$.
Consequently, the entanglement entropy is approximately obtained from a trimmed network $\rho_A$
shown in Fig.~\ref{fig.RDM_HOTRG_3}.

\begin{figure}
    \centering
    \begin{minipage}[c]{0.45\linewidth}
        \centering
        \includegraphics[keepaspectratio, height=4cm]{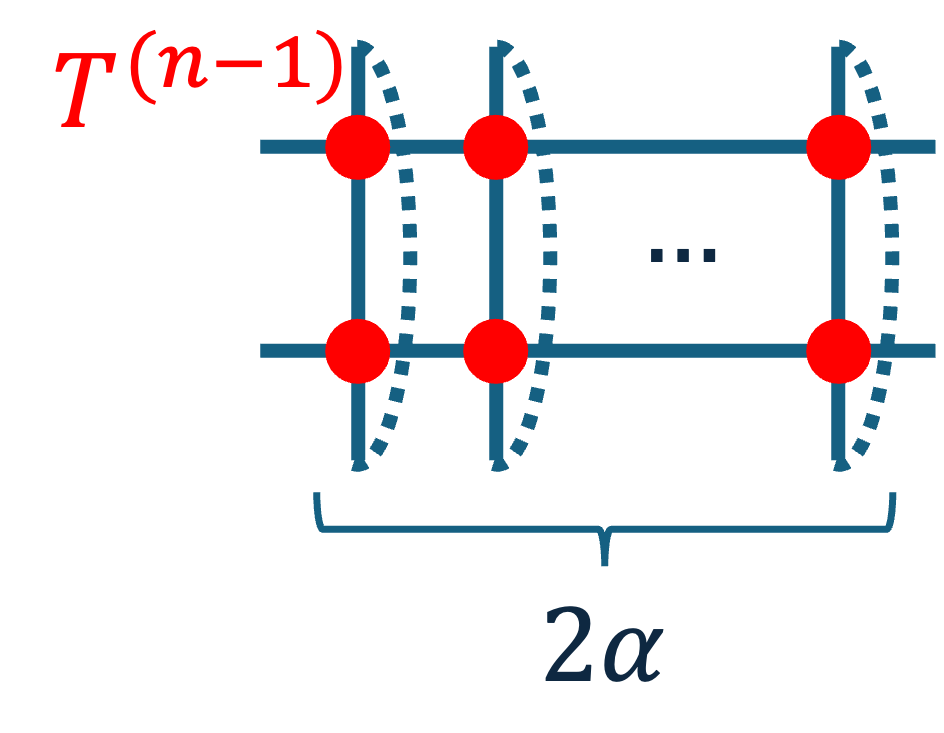}
        \subcaption[t]{Core matrix $C$.}
        \label{fig.Core}
    \end{minipage}
    \begin{minipage}[c]{0.45\linewidth}
        \centering
        \includegraphics[keepaspectratio, height=4cm]{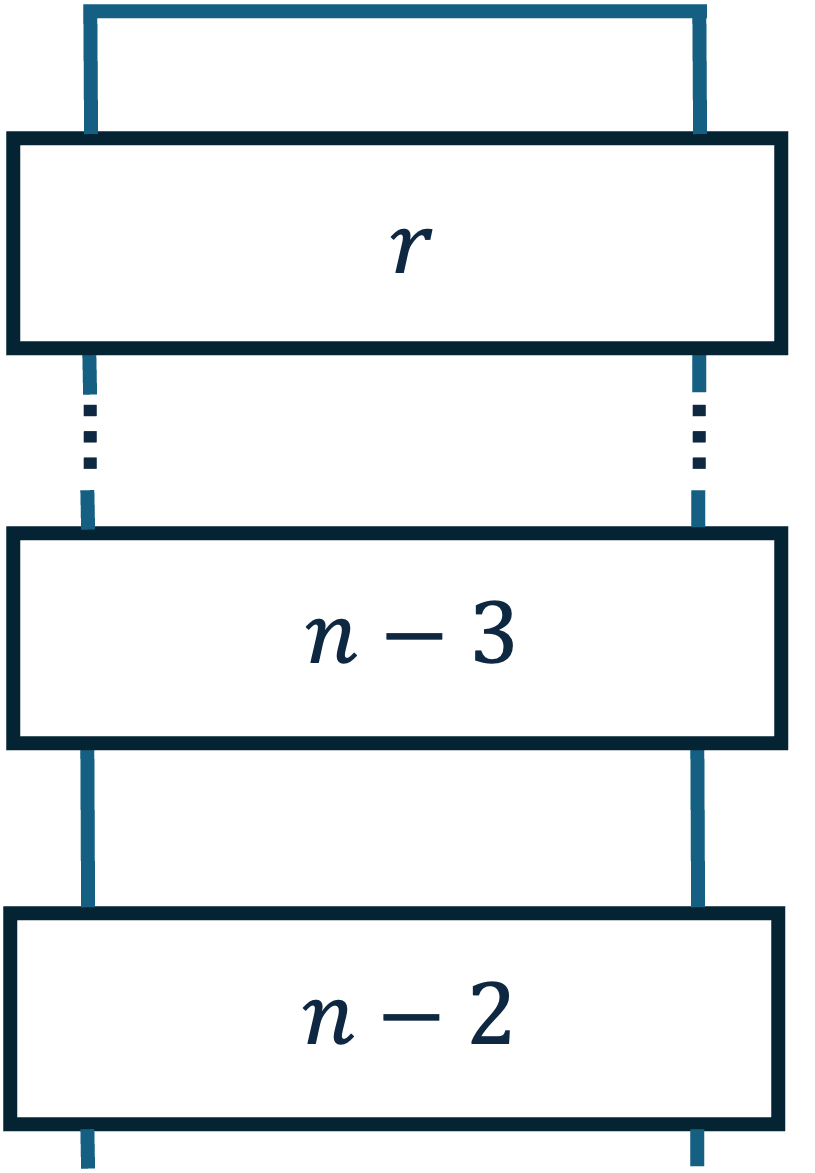}
        \subcaption[t]{Boundary factor $B$.}
        \label{fig.Boundary}
    \end{minipage}
    \caption{Core matrix $C$ and boundary factor $B$.}
\end{figure}

We give an algorithm that generalizes this method to arbitrary subsystem size $l$.
In the following discussion, we take $L=2^n$ and $N=\alpha \cdot 2^n$ ($\alpha \in \mathbb{N}$). 
It is convenient to use the binary representations of $l$ and $l-1$ for the generalized algorithm: 
\begin{subequations}
\begin{align}
    l =& \sum_{i=0}^{n-1} 2^i a_i \hspace{20pt} (a_i = 0, 1), \\
    l-1 =& \sum_{i=0}^{n-1} 2^i b_i \hspace{20pt} (b_i = 0, 1).
\end{align}
\end{subequations}
The trimmed tensor $\rho_A$ is made of two elements: the core matrix $C$, which consists of coarse-grained tensor $T^{(n-1)}$,
and the boundary factor $B$, which consists of isometry tensors $U^{(r)}, U^{(r+1)}, \dots, U^{(n-2)}$,
where $r$ is the largest integer that satisfies $a_r\neq b_r$.
The core matrix is represented by the network shown in Fig.~\ref{fig.Core}, 
and is a coarse-grained reduced density matrix of the half-space subsystem.

The boundary factor $B$ is created by combining isometries as shown in Fig.~\ref{fig.Boundary}.
The blank labeled with $k$ has a structure that changes with the value of $b_k$, and is filled with tensor shown in Fig. 5.
For example, in the case of $L=16$ and $l=5$, we have $a_3a_2a_1a_0=0101, \ b_3b_2b_1b_0=0100$
and $r=0$. Fig.~\ref{fig.BOUNDARY} shows the boundary factor of this case.
\begin{figure}
    \centering
    \begin{minipage}[t]{0.48\linewidth}
        \centering
        \includegraphics[keepaspectratio, scale=0.5]{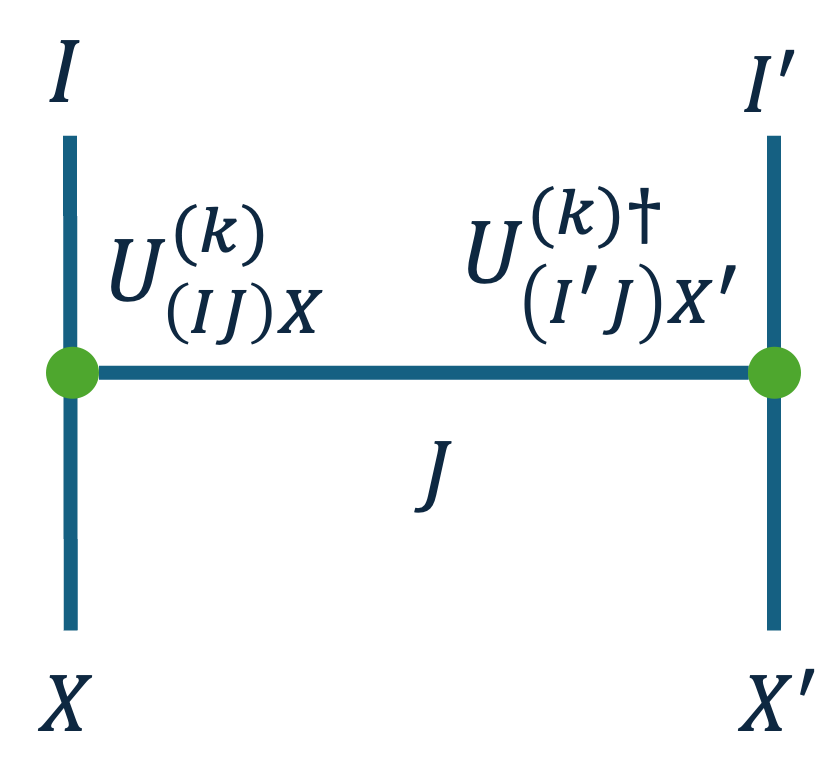}
        \subcaption{Case of $b_k = 0$.}
        \label{fig.ISO_CONT}
    \end{minipage}
    \centering
    \begin{minipage}[t]{0.48\linewidth}
        \centering
        \includegraphics[keepaspectratio, scale=0.5]{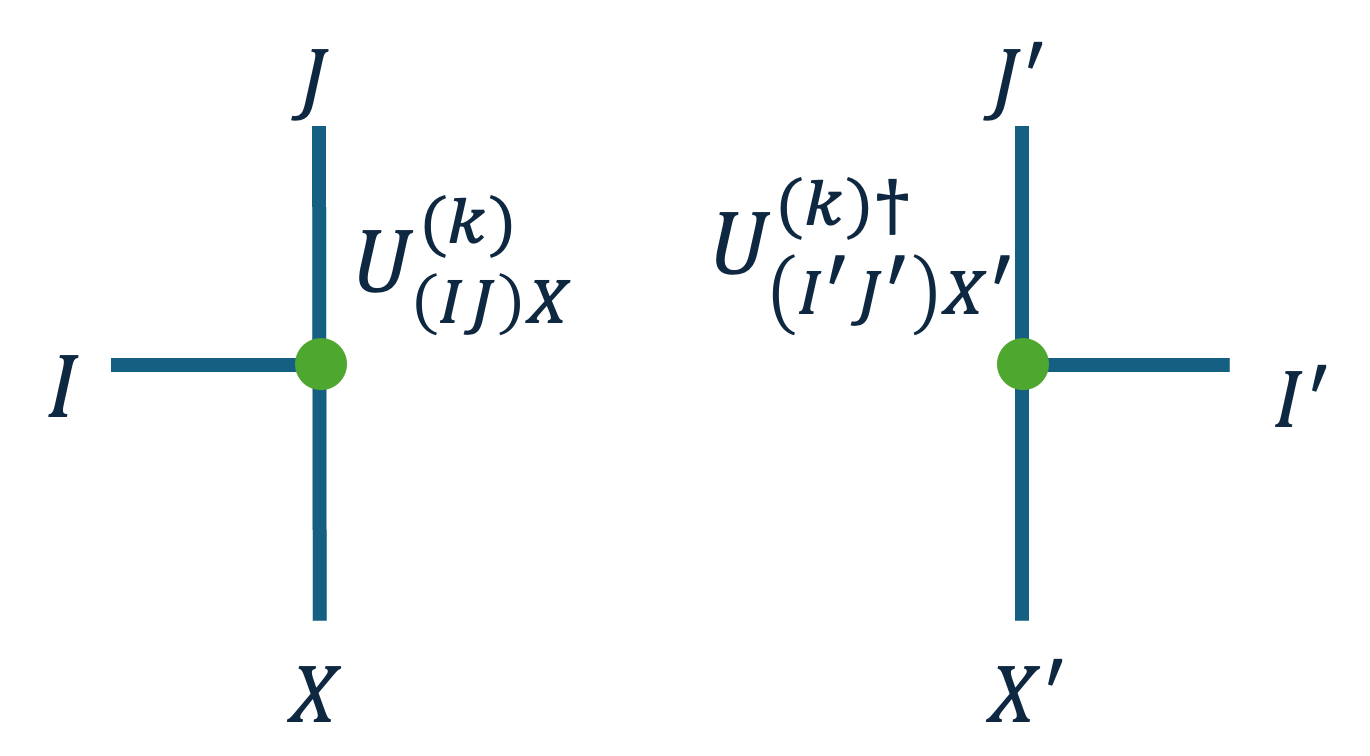}
        \subcaption{Case of $b_k = 1$.}
        \label{fig.ISO_PROD}
    \end{minipage}
    \caption{Structure of the box specified by $b_k$.}
    \label{fig.contraction}
\end{figure}

\begin{figure}[ht]
    \begin{minipage}[t]{0.4\linewidth}
      \centering  
      \includegraphics[scale=0.5]{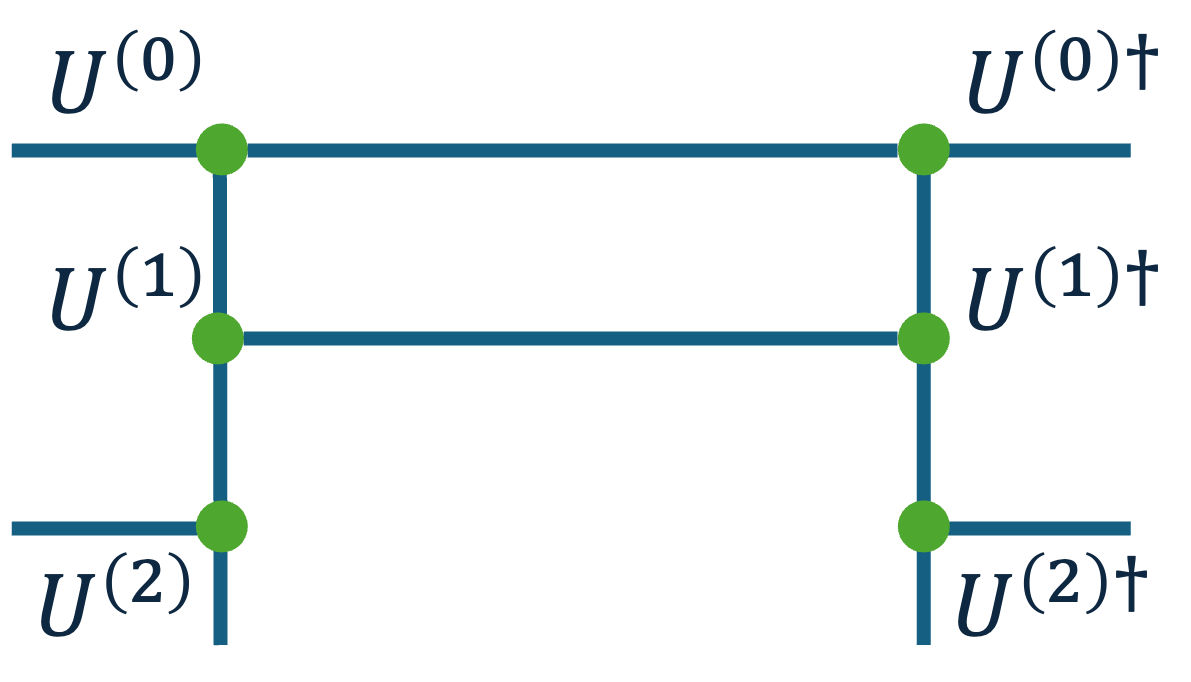}
        \caption{Boundary factor for $L=16$ and $l=5$.}
        \label{fig.BOUNDARY}
    \end{minipage}
    \hspace{0.05\linewidth}
    \nextfloat
    \begin{minipage}[t]{0.55\linewidth}
        \centering
        \begin{minipage}[b]{0.33\linewidth}
            \centering
            \includegraphics[keepaspectratio, scale=0.5]{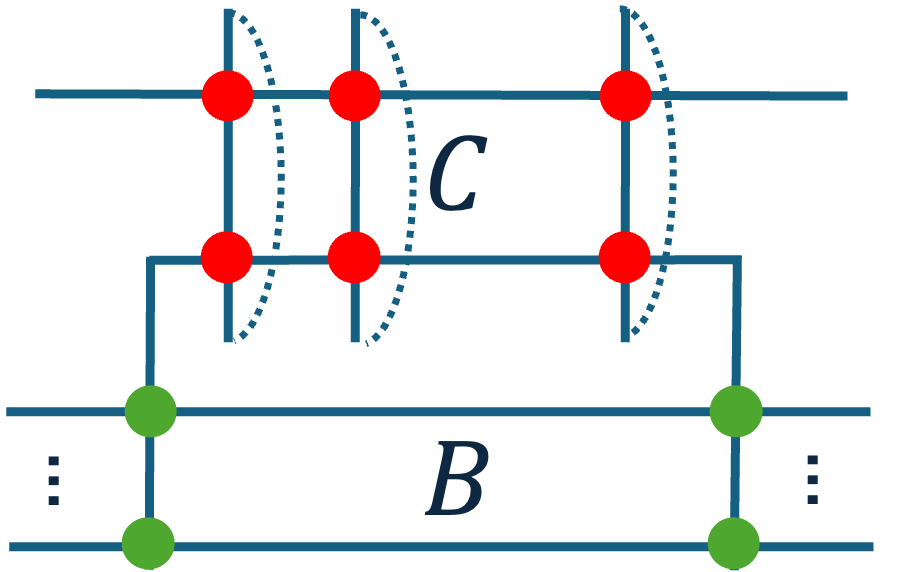}
            \subcaption{$l>L/2$}
            \label{fig.CORE_BOUNDARY_1}
        \end{minipage}
        \hspace{0.10\linewidth}
        \begin{minipage}[b]{0.33\linewidth}
            \centering
            \includegraphics[keepaspectratio, scale=0.5]{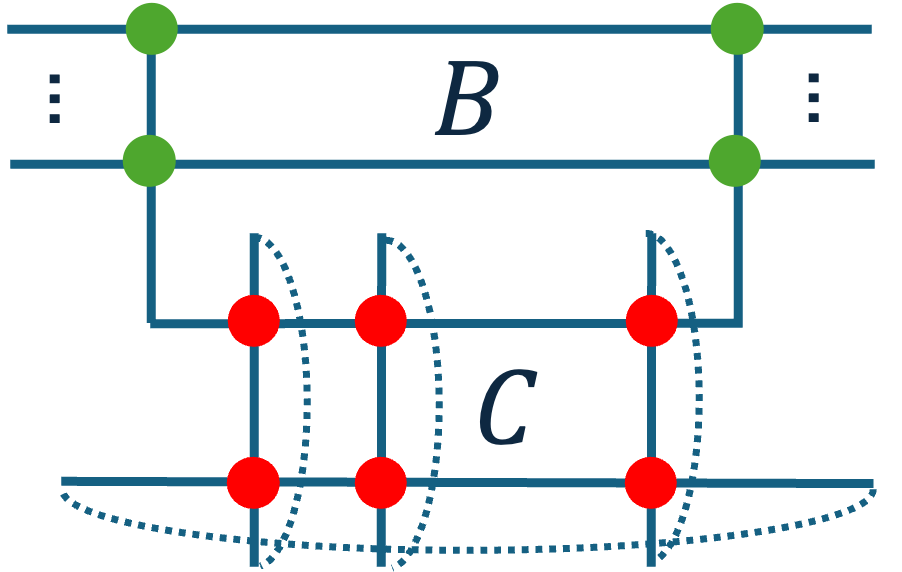}
            \subcaption{$l\leq L/2$}
            \label{fig.CORE_BOUNDARY_2}
        \end{minipage}
        \caption{Contraction of the boundary factor $B$ and the core matrix $C$.}
    \end{minipage}
\end{figure}
Finally, we contract the core matrix $C$ and the boundary factor $B$ to obtain $\rho_A$.
These tensors are contracted as Fig.~\ref{fig.CORE_BOUNDARY_1} for $l>L/2$
and Fig.~\ref{fig.CORE_BOUNDARY_2} for $l \le L/2$, respectively.
The computational cost for evaluating the EE, excluding the coarse-graining process,
is given by $O(D^{3h})$, where $h=\sum_i a_i$ represents the Hamming weight of subsystem size $l$ expressed in binary form.

\section{Numerical tests}
We present numerical results for the entanglement entropy of the one-dimensional quantum Ising model at the critical point.
We take $L=2^{10}=1024$  with $N=\alpha \cdot L \ (\alpha \in \mathbb{N})$
and calculate the entanglement entropy $S_A$ for several subsystem size $l$.
The temperature of the classical system is set to the critical temperature $K_c = 2/\log(1+\sqrt{2})$.
To keep the computational cost low, 
we limit our calculations to the case of Hamming weight $h \le 2$ taking
$l=2^m+q$ with $q=0,2^0,2^1,2^2, \dots, 2^{m-1}$ for $m=0,1, \dots, 9$.

\begin{figure}[htbp]
    \centering
  \begin{minipage}[t]{0.48\linewidth}
    \centering
    \includegraphics[scale=0.38]{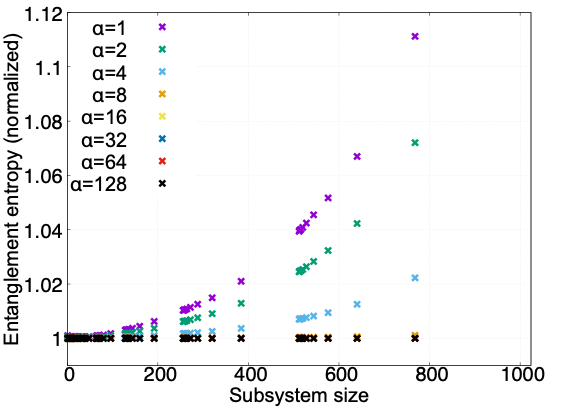}
    \caption{$S_A$ as a function of subsystem size for various $\alpha$ with $D_\text{cut}=64$.
        All results are normalized by dividing them by the data at $\alpha = 1024$.}
    \label{fig.alpha}
  \end{minipage}
  \hfil
  \begin{minipage}[t]{0.48\linewidth}
    \centering
    \includegraphics[scale=0.38]{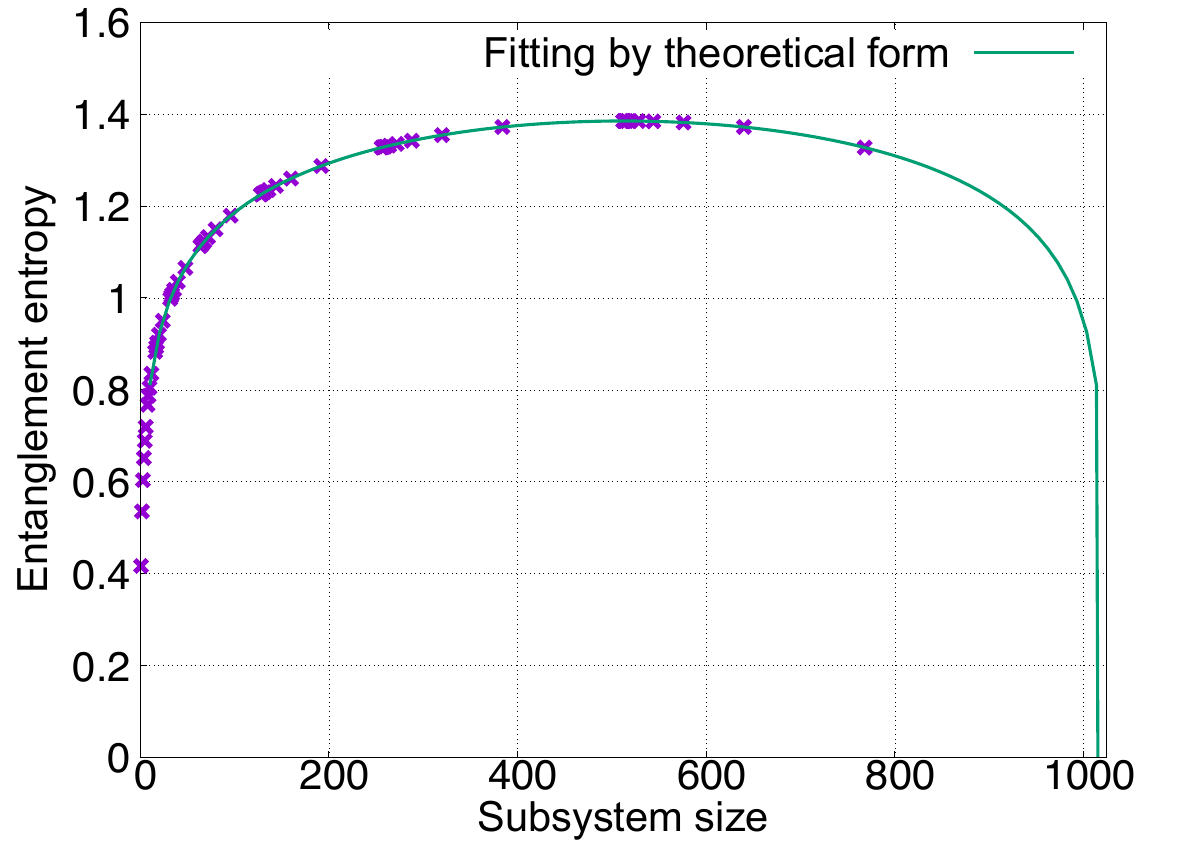}
    \caption{Subsystem size dependence of the entanglement entropy with $\alpha=16$ and $D_\mathrm{cut}=96$.}
    \label{fig.EE}
  \end{minipage}
\end{figure}

Fig.~\ref{fig.alpha} shows the $\alpha$-dependence of the entanglement entropy with $\alpha = 1,2,4, \dots, 128$.
The results in Fig.~\ref{fig.alpha} indicate that the entanglement entropy converges for $\alpha \geq 16$.
We thus set $\alpha=16$ in the following analysis, which is effectively regarded as the limit $N\to\infty$ in \eqref{eq.density_matrix_TN}.
Together with the finite temperature of the classical system,
this setting corresponds to the zero-temperature limit $\beta\to\infty$ of the one-dimensional quantum Ising model.

Fig.~\ref{fig.EE} shows the dependence of $S_A$ on the subsystem size $l$, along with the theoretical functional form
\begin{align}
    S_A(l,L) = \frac{c}{3} \log \left( L \sin \left( \frac{l}{L}\pi \right) \right) + k, \label{eq.EE_theoretical}
\end{align}
where $c$ is the central charge of the theory and $k$ is a constant, and they are determined by the fitting process below.
The expression \eqref{eq.EE_theoretical} corresponds to a one-dimensional,
finite-size quantum system at zero temperature mapped into a cylinder of spatial length $L$.
Our numerical results are plotted by purple crosses.
The errors in our calculation are so small that error bars are indistinguishable in Fig.~\ref{fig.alpha} and Fig.~\ref{fig.EE}.
The theoretical form accurately describes the data in the whole region, and the symmetry property $(S_A=S_{\bar{A}})$ is observed.

\begin{figure}
    \centering
    \begin{minipage}[b]{0.6\textwidth}
        \centering
        \includegraphics[scale=0.37]{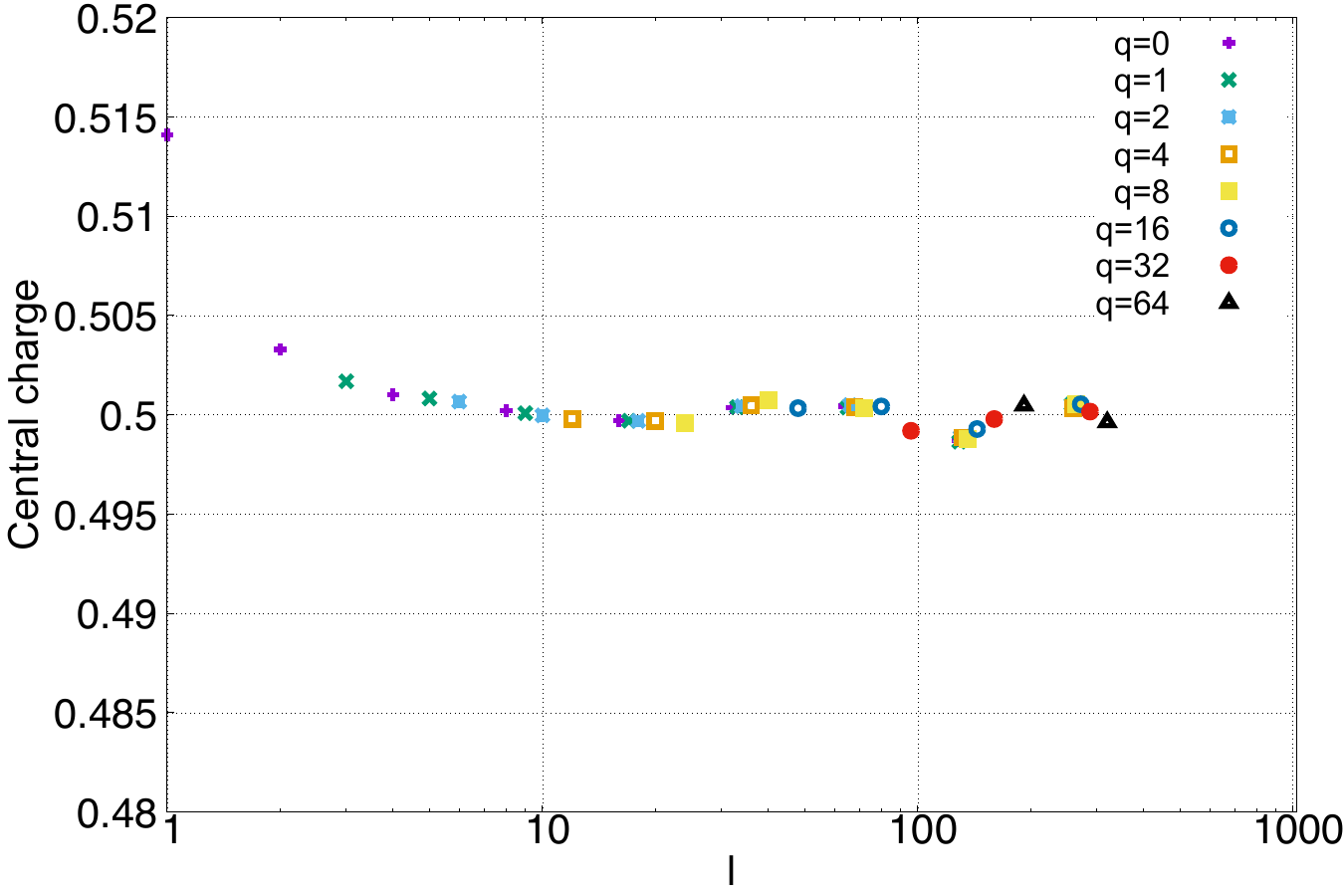}
        \caption{Central charge $c(l)$ computed by \eqref{eq.lrange}.}
        \label{fig.lrange}
    \end{minipage}
    \begin{minipage}[b]{0.38\textwidth}
        \centering
        \begin{tabular}{cl}
    	\hline	\hline	
    	$D_\mathrm{cut}$ & central charge	\\	\hline
    	$64$	&0.4998(2)	\\
    	$80$	&0.4999(1)	\\
    	$96$	&0.49997(8)	\\	\hline\hline	
        \end{tabular}
        \vspace{10pt}
        \captionof{table}{$D_{\rm cut}$-dependence of the central charge extracted from the entanglement entropy at the critical point.}
        \label{CCforDcut}
    \end{minipage}
\end{figure}

The central charge $c$ of the system is determined by fitting the result to the theoretical form.
To determine the fitting range,
we compute an effective central charge using the data of $S_A$ and the theoretical form \eqref{eq.EE_theoretical}:
\begin{align}
    c(l) = \frac{3(S_A(l',L) - S_A(l,L))}{\log \sin \frac{l'\pi}{L} - \log \sin \frac{l\pi}{L}}
    \label{eq.lrange}
\end{align}
with $l=2^m + q$ and $l'=2^{m+1} + q$.
The numerical result of $c(l)$ is shown in Fig.~\ref{fig.lrange}.
For small $l$, the results exhibit a visible $l$-dependence that is inconsistent with the expected constant behavior.
This observation suggests that the data for small $l$ should be excluded in the fitting process.
Consequently, we choose $7 \leq l \leq 768$ as the fit range.
We then perform a fitting using the functional form in \eqref{eq.EE_theoretical}, with two parameters $c$ and $k$.
To estimate the error in the central charge, 
we first solve \eqref{eq.EE_theoretical} with respect to $c$ for each $7\le l \le 768$.
In this process, $k$ is set to the value obtained through the previous fitting process,
and $S_A(l,L)$ is set to our numerical result.
For all $7\le l \le 768$, we calculate the difference between the central charge obtained from this procedure
and the one determined through the fitting.
The maximum difference among all $l$ values is taken as the error.
The final result is given as
\begin{align}
	c=0.49997(8) \hspace{20pt} \text{for} \hspace{8pt} D_{\rm cut}=96 \ .
\end{align}
The same analysis is repeated for other bond dimensions $D_\mathrm{cut} = 64$ and $80$, and
the results are summarized in Table \ref{CCforDcut}.

\section{Summary and outlook}
In this paper, we presented a method for computing the entanglement entropy of subsystems of any size.
We applied this method to the one-dimensional quantum Ising model, representing the density matrix as a (1+1)-dimensional tensor network.
As a demonstration, we calculated the dependence of the entanglement entropy on the subsystem size and confirmed that it exhibits the expected logarithmic scaling.
Additionally, we accurately extracted the central charge from numerical data of the entanglement entropy.
These results support the validity of our method.

In future work, we plan to extend our method to more complex models,
including higher-dimensional ($d\geq3$) theories, and investigate their
phase transitions using entanglement entropy as a probe.
The Ryu-Takayanagi formula \cite{Ryu:2006bv} implies that one can extract the geometry of spacetime from entanglement entropy.
We hope that our method can serve as a useful tool in advancing this direction.

\acknowledgments
S.T. was supported by JSPS KAKENHI Grant Numbers 21K03531, and 22H05251.
D.K. and G.T. were supported by JSPS KAKENHI Grant Numbers 21K03537, 22H01222, 23K22493, and 23H00112.
This work was supported by JST SPRING, Grant Number JPMJSP2135.
Useful discussions with the participants of LATTICE2024 are acknowledged.

\bibliographystyle{JHEP}
\bibliography{ref}

\end{document}